\documentstyle[psfig]{caosp}
\begin{document}
\def\teff{$T_{\rm eff}$}
\def\lgg{$\log\,{g}$}
\def\vt{$\xi_{\rm t}$}
\def\vsini{$v\cdot \sin i$}
\def\kms{\,km\,s$^{-1}$}
\def\and{$\alpha$ And}
\def\kcnc{$\kappa$ Cnc}
\def\xlup{$\xi$ Lup}
\def\icrb{$\iota$ CrB}
\htitle{Abundance analysis of SB2 binary stars}
\hauthor{T. Ryabchikova}
\title{Abundance analysis of SB2 binary stars with HgMn primaries}
\author{T. Ryabchikova \inst{1}}
\institute{Institute of Astronomy, RAS, Moscow, Russia}
%
\maketitle

\begin{abstract}
We present a short review of the abundances in the atmospheres of SB2
systems with Mercury-Manganese (HgMn) primaries.
Up to now a careful study has been made for both components of 8 out of 17
known
SB2 binaries with orbital periods shorter than 100 days and mass
ratio ranging from 1.08 to 2.2. For all eight systems we observe a lower Mn
abundance in the secondary's atmospheres than in the primary's. Significant
difference in the abundances is also found for some peculiar elements such
as Ga, Xe, Pt. All secondary stars with effective temperatures less than
10000 K show abundance characteristics typical of the metallic-line stars.
\end{abstract}

\section{Introduction}
The study of abundances in atmospheres of SB2 systems with peculiar components
may provide constraints on the origin of the abundance anomalies. It is
natural to believe that both components in a binary system with a period
less than 100 days form from the same protostellar cloud, therefore their
initial abundances have to be identical. Any observed difference and its
dependence on the mass and/or atmospheric parameters of the star may show us
the development of abundance anomalies during stellar evolution.
Among peculiar stars, the highest frequency of binary systems is observed for
the non-magnetic HgMn and metallic-line (Am) stars. We give here a short review
of the atmospheric abundances in SB2 binary stars with HgMn primaries.
\section{The main characteristics of SB2 binary stars}
                                
There are 15 SB2 systems in Lebedev's catalogue (Lebedev 1987) whose primaries
belong to the HgMn group. For two other systems, $\alpha$ And and $\kappa$ Cnc,
orbital elements are presented in poster papers of this conference. An
abundance analysis of both components is carried out for eight SB2 HgMn binary
stars. Their main orbital parameters are presented in Table 1, while
atmospheric
parameters, masses and radii are given in Table 2. Masses and radii were
estimated using mass ratios and evolutionary tracks by Schaller et al. (1992).

\begin{table}
\caption{Orbital parameters of SB2 HgMn stars}
\label{table1}
\begin{flushleft}
\begin{small}
\begin{tabular}{lcccccccc}
\noalign{\smallskip}
\hline
      &\and&\kcnc&112 Her&46 Dra&\icrb&AR Aur&HR 4072&\xlup \\
\hline
P(days)&97 &6.39&6.36  &9.81 &35.5&4.13 &11.6  &15.3 \\
e     &0.53&0.14&0.12  &0.20 &0.56&0.01 &0.26  &0.00 \\
i     &105&81  &16    &17&3& 88&40&72 \\
$m_{A}/m_{B}$& 2.0&2.2&1.98&1.12 &1.5 &1.14 &1.67  &1.42 \\
\hline
\end{tabular}
\end{small}
\end{flushleft}
\end{table}

\begin{table}
\caption{Spectroscopic parameters, masses, and radii of SB2 HgMn stars}
\label{table2}
\begin{flushleft}
\begin{small}
\begin{tabular}{rrcrlll}
\noalign{\smallskip}
\hline
 Star  &\teff&\lgg&\vsini&M(M$_\odot$)& R(R$_\odot$)& Synchron.\\
\hline
\and\, A &13800&3.86&51.~  &3.8  &2.7 &no\\
       B&8500&4.20&110.~ &1.85 &1.65&no\\
\kcnc\, A&13200&3.81&  6.~ &4.5  &5.0 &subsynch.\\
        B& 8500&4.00& 40.~ &2.1  &2.4 &subsynch.\\
112 Her A&13100&4.21&  6.~ &3.9  &3.3 &yes \\
        B& 8500&4.20&  8.~ &2.0  &2.1 &yes\\
46 Dra A &11700&4.11&  5.~ &3.3  &3.3 &yes\\
       B &11100&4.11&  5.~ &2.9  &2.7 &yes\\
\icrb\, A&11250&3.75  &$\le$0.5&3.5  &3.8 &subsynch.\\
        B& 9250&4.00  &$\le$0.5&2.3  &2.3 &subsynch.\\
AR Aur A &10950&4.33& 23.~ &2.5  &1.78&yes\\
       B &10350&4.28& 23.~ &2.3  &1.82&yes\\
HR4072 A &10900&4.07  &$\le$2.0&2.8  &2.4 &subsynch.\\
       B&8900&4.20  &$\le$2.0&1.7  &1.7 &subsynch.\\
\xlup\, A&10650&3.90&  0.~ &3.0  &3.2 &subsynch.\\
        B& 9200&4.20&  2.~ &2.1  &2.0 &subsynch.\\
\hline
\end{tabular}
\end{small}
\end{flushleft}
\end{table}

The positions of the stars on a \lgg - \teff\, plot are shown in Fig. 1 with
evolutionary tracks overlaid. Filled circles refer to the primaries,
the size of the circle being proportional to the \teff\, of the primary.

\begin{figure}[hbt]
\psfig{figure=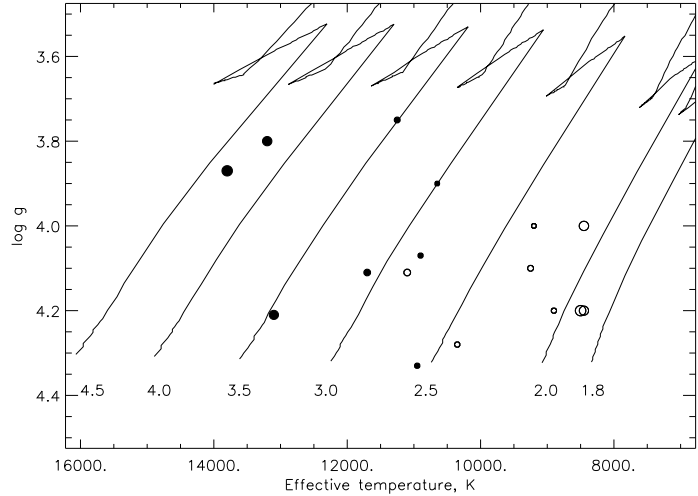,width=10cm}
\caption[]{Position of the components of SB2 HgMn primaries (filled circles)
           and secondaries (open circles) on the \lgg -- \teff\, plot.
           Evolutionary tracks are shown by solid lines. The size of the circle

           is proportional to the \teff\, of the primary}
\label{evolut}
\end{figure}

\section{Chemical abundances of the components}

The reader can find a complete set of abundances in the atmospheres of
the SB2 components in the following papers: $\iota$ CrB A, B (Adelman, 1989),
HR 4072 A, B (Adelman, 1994), $\chi$ Lup A, B (Wahlgren et al., 1994),
112 Her A, B (Ryabchikova et al., 1996), and 46 Dra A, B
(Adelman et al., 1998).
For AR Aur we used equivalent widths and atmospheric parameters
from Khokhlova et al. (1995), and recalculated abundances with WIDTH9.
For $\alpha$ And A, B and for $\kappa$ Cnc B we used results of the
preliminary abundance analysis presented in this conference.

Table 3 gives the abundances of some elements from He to Ba and Pt-Au-Hg in
the atmospheres
of the primaries, while the abundance data for the secondaries are shown
in Table 4. As a comparison, we give abundances in the hot Am star $\theta$
Leo with \teff=9250 K (Adelman, 1988) in the last column of Table 4.
Data on Ga abundance for $\iota$ CrB A, HR 4072 A, and $\xi$ Lup A are taken
from Smith (1996), and those on Hg in HR 4072 A, $\xi$ Lup A, and $\theta$ Leo
are taken from Smith (1997). The platinum abundance was recalculated using
gf-values from Dworetsky et al. (1984),
and the gold abundance in the optical
region was recalculated with the real (not assumed) gf-values from Rosberg
\& Wyart (1997).

\begin{table*}
\label{table3}
\caption{Abundances in the atmospheres of SB2 primaries  ($\log$ N/H)}
\begin{scriptsize}
\begin{tabular}{lccccccccccccccccc}
\noalign{\smallskip}

Species&$\alpha$ AndA&$\kappa$ Cnc A&112HerA&46 DraA&$\iota$ CrB A&AR AurA
&HR4072A & $\chi$ LupA&Sun\\
  He I&-2.00~&-2.00&-2.43~&-2.00&~-1.82& ...   &-1.46~ &~-1.68~ &~(-1.01) \\
  C II&-3.88~&-3.97&-5.36:&-3.92&~-4.07 &~-2.72~ &-3.23~ &~-3.46~ &~~-3.43~ \\
  O  I& ... &-3.23&-3.36:&-3.71&  ...& ...  &...   &~-3.48~ &~~-3.14~ \\
  Mg II&-4.98~&-5.00&-5.56~&-4.70&~-4.84 &~-4.46~ &-4.57~ &~-4.55~ &~~-4.42~ \\
  Al I& ...&...& ...&...& ...  &~-6.15~ &-6.42~ &~-6.18~ &~~-5.53~ \\
  Al II& ...&...&-6.42:& ...& ...& ...   &-6.85~& ...   &~~-5.53~ \\
  Si II&-4.38~&-4.48&-5.05~&-4.82&~-4.51 &~-4.24~ &-4.53~ &~-4.33~ &~~-4.45~ \\
  P II&-4.80~&-4.73&-4.71~&-5.11&~-5.69& ...   &-5.54~ &~-5.68~ &~~-6.55~ \\
  P III&-4.83~&-4.77&-4.80~& ...& ...& ...  &...&...   &~~-6.55~ \\
  S II&-5.48~&-5.56&-5.84~& ... &~-4.79 &$<$-4.70~&-4.64~ &~-4.36~ &~~-4.67~ \\
  Ca I& ...&...&-4.55:& ...& ...& ...   &-5.24~ &~-5.98~ &~~-5.64~ \\
  Ca II&-5.00~&-5.67& ... &-5.49&~-5.07 &~-6.14~ &-4.80~ &~-6.78~ &~~-5.64~ \\
  Sc II& ... &-8.37&-9.06:&-8.89&  ...  &$<$-8.50~&-8.40~ &-10.28~ &~~-8.90~ \\
  Ti II&-6.70~&-6.82&-6.30~&-6.73&~-6.72 &~-6.26~ &-6.16~ &~-6.54~ &~~-7.01~ \\
  V II& ... &-7.60& ... &-7.93&~-7.56 &~-7.76~ &-8.68~ &~-8.13~ &~~-8.00~ \\
  Cr II&-6.00~&-6.42&-6.50~&-6.38&~-6.01 &~-5.82~ &-5.62~ &~-6.18~ &~~-6.26~ \\
  Mn I& ... &-4.39&-4.74~&-5.33&~-5.14& ...   &-5.51~ &~-6.78~ &~~-6.45~ \\
  Mn II&-3.80~&-4.10&-4.91~&-5.14&~-5.11 &~-5.08~ &-5.38~ &~-6.38~ &~~-6.45~ \\
  Fe I&-4.13~&-4.49&-3.55~&-4.08&  ...  &~-3.86~ &-4.00~ &~-4.30~ &~~-4.52~ \\
  Fe II&-4.18~&-4.57&-3.60~&-4.15&~-4.44 &~-3.82~ &-4.08~ &~-4.34~ &~~-4.52~ \\
  Fe III&  ... &-4.44&-3.66~&-3.91&~-4.32& ...   &-3.97~& ...   &~~-4.52~ \\
  Ni II& ... &-6.18&-6.07~&-6.29&~-6.62 &~-6.93~ &-6.62~ &~-6.08~ &~~-5.75~ \\
  Ga II&-4.63~&-4.75&-5.27~&-5.21&~-7.60& ...   &-7.30  &~-7.35~ &~~-9.12~ \\
  Sr II&-8.08~&-8.54&-8.57:&-8.00&~-7.20 &~-6.79~ &-6.49~ &~-7.03~ &~~-9.10~ \\
  Y II&-8.20:&-8.33&-8.04:&-7.91&~-7.41 &~-7.14~ &-6.56~ &~-8.08~ &~~-9.76~ \\
  Zr II&-7.50:& ...&-7.74~&-7.58&~-7.52 &~-7.59~ &-8.04~ &~-8.88~ &~~-9.40~ \\
  Xe II&-4.88~&-4.75&-5.78~&-5.68&  ...& ...  &...&... &~(-9.77)\\
  Ba II& ...&...& ... &-9.05  &$<$-8.42&~-8.97~ &-9.14~ &~-8.78~ &~~-9.87~ \\
  Pt II& ...&...& ... &-6.93&~-6.80 &~-4.97~ &-5.91~ &~-6.23~ &~-10.2~~ \\
  Au II& ...&...& ... &-6.80&  ...& ...   &-6.64: &~-6.69~ &~-10.99~ \\
  Hg I& ...&...& ...&... &~-6.02& ...   &-5.71~ &~-5.86~ &(-10.91) \\
  Hg II&-6.00:&-6.00&-6.00~&-5.80&~-6.01 &~-6.25~ &-5.28~ &~-5.60~ &(-10.91) \\
\end{tabular}
\end{scriptsize}
\end{table*}

\begin{table*}
\label{table4}
\caption{Abundances in the atmospheres of SB2 secondaries ($\log$ N/H)}
\begin{scriptsize}
\begin{tabular}{lccccccccccccccccc}
\noalign{\smallskip}

 Species    &46 DraB&AR AurB&$\iota$ CrB&$\chi$ LupB& HR4072B& 112HerB
&$\alpha$ And B&$\kappa$ Cnc B&$\theta$ Leo\\
  He I&  -1.78&  ...&  ...&  ...&  ...&  ...&  ...&  ...& ~-1.22 \\
  Mg I&  -4.63&  ...&~-4.88&~-4.66&~-5.12&~-5.04~&   ...&  ...& ~-4.53 \\
  Mg II&  -4.58& -4.35&  ...&~-4.66&~-4.64&~-5.25~& ~-4.42~& -4.42& ~-4.66 \\
  Si I&   ...&  ...&  ...&~-5.09&~-5.37&~-5.38:&   ...&  ...&   ...  \\
  Si II&  -4.55& -4.21&  ...&~-4.56&~-5.18&  ...&~-3.91~& -4.45& ~-4.46 \\
  Ca I&  -5.47&  ...&  ...&~-6.28&~-6.45&~-5.83~& ~-6.26~& -5.70& ~-5.76 \\
  Ca II&  -5.55& -6.28&  ...&  ...&  ...&  ...&~-5.96~& -5.70& ~-5.57 \\
  Sc II&  -9.69&  ...&  ...&~-9.67&-10.07&~-9.36~&   ...&  ...& ~-9.27 \\
  Ti II&  -6.48& -6.67&  ...&~-6.93&~-7.06&~-6.79~& ~-6.90~&   ...& ~-6.95 \\
  Cr I&  -5.88&  ...&~-6.32&~-6.29&~-6.19&~-5.98~& ~-5.96~&   ...& ~-6.31 \\
  Cr II&  -5.95& -5.79&  ...&~-6.09&~-6.16&~-5.94~& ~-5.96~&   ...& ~-6.32 \\
  Mn I&  -5.68&  ...&  ...&~-6.70&~-6.51&~-5.97~& ~-6.10~&   ...& ~-6.70 \\
  Mn II&  -6.11& -5.67&  ...&~-6.63&~-5.97&~-6.00:&   ...&  ...& ~-6.31 \\
  Fe I&  -4.09& -4.03&~-4.41&~-4.47&~-4.51&~-4.44~& ~-4.20~& -4.20& ~-4.52 \\
  Fe II&  -4.20& -4.05&  ...&~-4.51&~-4.54&~-4.45~& ~-4.30~& -4.20& ~-4.43 \\
  Co I&   ...&  ...&  ...&  ...&~-7.09&~-6.42~&   ...&  ...& ~-6.95 \\
  Ni I&   ...&  ...&  ...&  ...&~-5.46&~-5.18~&   ...&  ...& ~-5.35 \\
  Ni II&   ...& -5.16&  ...&~-5.42&~-5.45&~-5.05~&$\leq$-5.50~&...& ~-5.34\\
  Ga II &$\leq$-7.20&...&  ...&  ...&  ...&  ...&  ...& ...& ...\\
  Ba II&  -9.06& -8.05&  ...&~-7.67&~-7.45&~-8.08~& ~-8.75~& -8.70& ~-8.50  \\
  La II&   ...&  ...&  ...&  ...&-10.13&~-9.73~&   ...&  ...& ~-9.88 \\
  Ce II&   ...&  ...&  ...&  ...&~-9.37&~-8.62~&   ...&  ...& ~-9.05 \\
  Pr II&   ...&  ...&  ...&  ...&  ...&~-9.08~&   ...&  ...&   ...  \\
  Nd II&   ...&  ...&  ...&  ...&~-9.62&~-8.81~&   ...&  ...& ~-9.62 \\
  Sm II&   ...&  ...&  ...&  ...&  ...&~-8.98~&   ...&  ...&   ...  \\
  Eu II&   ...&  ...&  ...&-10.49&  ...&-10.75~&   ...&  ...& -10.45 \\
  Gd II&   ...&  ...&  ...&  ...&  ...&~-9.34~&   ...&  ...& ~-9.84 \\
  Pt II&  -5.72& -5.74&  ...&  ...&  ...&  ...&  ...&  ...& ... \\
  Au II&  -6.33&  ...&  ...&  ...&  ...&  ...&  ...&  ...& ...\\
  Hg I&  -5.96&  ...&  ...&  ...&  ...&  ..  .&   ...&  ...& ...\\
  Hg II&  -5.50&  ...&  ...&  ...&  ...&  ...&  ...&  ...& ~-9.50\\
\end{tabular}
\end{scriptsize}
\end{table*}

\section{Conclusions}

\begin{itemize}
\item{Abundances in the atmospheres of SB2 systems may differ significantly
      even when both stars have practically equal masses}
\item{A comparison of the abundances in $\alpha$ And, $\kappa$ Cnc, and 112 Her
      shows that the observed peculiarities do not correlate with the orbital
periods or eccentricities}
\item{In three hot primaries of practically equal temperatures, a violation
      of the odd-even effect in Mn-Fe is observed in non-synchronized systems}
\item{The gallium abundance drops by about 1.5 dex within a narrow temperature
      range 11300-11600 K. It follows from the paper by Smith (1996) and it is
      supported by the abundances in 46 Dra. The only exception is HR 7775,
which has \teff=10650 K and a high gallium abundance}
\item{There is no correlation between Hg abundance in the primaries and
      their atmospheric parameters, masses or orbital parameters, which
      supports the same conclusion made by Smith (1997)}
\item{The detailed study of the atmospheres of HR 4072 B and 112 Her B shows
that their abundance pattern is similar to abundances in Am and roAp stars.
      The only feature which allows to classify both secondaries as Am stars
      is the Co/Ni ratio: [Co/Ni]$\approx$-1.0 in HR 4072 B, 112 Her B and in
Am stars, while [Co/Ni]$\approx$0.5 in all roAp stars with detailed abundance
analysis}
\item{In view of the similarity of the abundances of the main elements in the
      atmospheres of the secondary stars with \teff$<$10000 K, it seems
      to be possible to classify them as Am stars}
\end{itemize}

\acknowledgements
This work has been partially supported by a Grant 1.4.1.5 of the Russian
Federal
program ``Astronomy''.

\end{document}